\documentclass[12pt]{article}

\usepackage{amsmath}
\usepackage{amsfonts}
\usepackage{amsthm}
\usepackage{graphicx}
\usepackage{caption}
\usepackage{subcaption}
\usepackage{enumerate}
\usepackage{hyperref}
\usepackage[parfill]{parskip} 

\usepackage{fancyhdr}
\numberwithin{equation}{section}

\begin{document}

\title{Using Non-Linear Difference Equations to Study Quicksort Algorithms}
\author{Yukun Yao}

\maketitle

\begin{abstract}
Using non-linear difference equations, combined with symbolic computations, we make a detailed study of the running times of numerous variants of the celebrated Quicksort algorithms, where we consider the variants of single-pivot and multi-pivot Quicksort algorithms as discrete probability problems. With non-linear difference equations, recurrence relations and experimental mathematics techniques, explicit expressions for expectations, variances and even higher moments of their numbers of comparisons and swaps can be obtained. For some variants, Monte Carlo experiments are performed, the numerical results are demonstrated and the scaled limiting distribution is also discussed.

\bigskip


\end{abstract}
\leavevmode
\\
\\
\\

\section{Introduction}

A sorting algorithm is an algorithm that rearranges elements of a list in a certain order, the most frequently used orders being numerical order and lexicographical order. Sorting algorithms play a significant role in computer science since efficient sorting is important for optimizing the efficiency of other algorithms which require input data to be in sorted lists. In this paper, our focus is {\it Quicksort}. 

Quicksort was developed by British computer scientist Tony Hoare in 1959 and published in 1961. It has been a commonly used algorithm for sorting since then and is still widely used in industry. 

The main idea for Quicksort is that we choose a pivot randomly and then compare the other elements with the pivot, smaller elements being placed on the left side of the pivot and larger elements on the right side of the pivot. Then we recursively apply the same operation to the sublists obtained from the partition step. As for the specific implementations, there can be numerous variants, some of which are at least interesting from a theoretical perspective despite their rare use in the real world.

It is well-known that the worst-case performance of Quicksort is $O(n^2)$ and the average performance is $O(n \log n)$. However, we are also interested in the explicit closed-form expressions for the moments of Quicksort's performance, i.e., running time, in terms of the number of comparisons and/or the number of swaps. In this paper, only lists or arrays containing distinct elements are considered.

The paper is organized as follows. In Section 2, we review related work on the number of comparisons of 1-pivot Quicksort, whose methodology is essential for further study. In Section 3, the numbers of swaps of several variants of 1-pivot Quicksort are considered. In Section 4, we extend our study to multi-pivot Quicksort. In Section 5, the technique to obtain more moments and the scaled limiting distribution are discussed. In the last section we discuss some potential improvements for Quicksort, summarize the main results of this paper and make final remarks on the methodology of experimental mathematics. 

{\bf Accompanying Maple Package} 

This  article is accompanied by Maple packages {\tt QuickSort.txt} and {\tt Findrec.txt} available from the front of this article

{\tt http://sites.math.rutgers.edu/\~{}yao/QuickSort.html} \quad.

{\tt QuickSort.txt} is the main package of this paper and all procedures mentioned in the paper are from this package unless noted otherwise. {\tt Findrec.txt} is mainly used to find a recurrence relation, i.e., difference equation of moments from the empirical data. 

\section{Related Work}
In the masterpiece of Shalosh B. Ekhad and Doron Zeilberger [3], they managed to find the explicit expressions for expectation, variance and higher moments of the number of comparisons of 1-pivot Quicksort with an experimental mathematics approach, which is also considered as some form of
``machine learning." Here we will review the results they discovered or rediscovered. 

Let $C_n$ be the random variable  ``number of comparisons in Quicksort applied to lists of length $n$," $n \geq 0$.

{\bf Theorem 2.1} ([12], p.8, end of section 1.3; [5], Eq. (2.14), p. 29, and other places)
$$
E[C_n]=2(n+1) H_n - 4n .
$$
Here $H_n$ are the {\it Harmonic numbers}
$$
H_n:=\sum_{i=1}^{n} \frac{1}{i} .
$$

In following theorems, we introduce the notation 
$$
H_k(n) := \sum_{i=1}^n \frac{1}{i^k}.
$$

{\bf Theorem 2.2} (Knuth, [11], answer to Ex. 8(b) in section  6.2.2))

$$
var[C_n] = n ( 7\,n+13 )  \, - \, 2\,(n+1)\, H_{{1}} ( n )  -4\, ( n+1 ) ^{2}H_{{2}} ( n )  .
$$

Its asymptotic expression is
$$
( 7 \,- \,\frac{2}{3}\,{\pi }^{2} ) {n}^{2}+ ( 13-2\,\ln  ( n
 ) -2\,\gamma-4/3\,{\pi }^{2} ) n-2\,\ln  ( n ) 
-2\,\gamma-2/3\,{\pi }^{2} \, + \, o(1)  .       
$$

{\bf Theorem 2.3} (Zeilberger, [3]) The third moment about the mean of $C_n$ is 
$$
-n ( 19\,{n}^{2}+81\,n+104 ) +H_{{1}} ( n ) 
 ( 14\,n+14 ) +12\, ( n+1 ) ^{2}H_{{2}} ( n
 ) +16\, ( n+1 ) ^{3}H_{{3}} ( n )  .
$$
It is asymptotic  to
$$
( -19+16\,\zeta  ( 3 )  ) {n}^{3}+ ( -81+2
\,{\pi }^{2}+48\,\zeta  ( 3 )  ) {n}^{2}+ ( -104+
14\,\ln  ( n ) 
+14\,\gamma+4\,{\pi }^{2}+48\,\zeta  ( 3
 )  ) n 
$$
$$
+14\,\ln  ( n ) +14\,\gamma+2\,{\pi }^{2}
+16\,\zeta  ( 3 ) \, + \, o(1)  .
$$
It follows that the limit of the scaled third moment (skewness) converges to
$$
{\frac{-19+16\,\zeta  ( 3 ) }{ ( 7-2/3\,{\pi }^{2} ) ^{3/2}}} \, = \, 0.8548818671325885 \dots \quad .
$$

{\bf Theorem 2.4} (Zeilberger, [3]) The fourth moment about the mean of $C_n$ is 
$$
\frac{1}{9} \,n ( 2260\,{n}^{3}+9658\,{n}^{2}+15497\,n+11357 ) -2\,
 ( n+1 )  ( 42\,{n}^{2}+78\,n+77 ) H_{{1}}
 ( n ) 
$$
$$
+12\, ( n+1 ) ^{2} ( H_{{1}} ( n
 )  ) ^{2}+ ( -4\, ( 42\,{n}^{2}+78\,n+31
 )  ( n+1 ) ^{2}+48\, ( n+1 ) ^{3}H_{{1}}
 ( n )  ) H_{{2}} ( n ) 
$$
$$
+48\, ( n+1
 ) ^{4} ( H_{{2}} ( n )  ) ^{2}-96\,
 ( n+1 ) ^{3}H_{{3}} ( n ) -96\, ( n+1
 ) ^{4}H_{{4}} ( n ) .
$$
It is asymptotic  to
$$
 ( {\frac {2260}{9}}-28\,{\pi }^{2}+{\frac {4}{15}}\,{\pi }^{4}
 ) {n}^{4}+ ( {\frac {9658}{9}}-84\,\ln  ( n ) -
84\,\gamma+1/6\, ( -648+48\,\ln  ( n ) +48\,\gamma
 ) {\pi }^{2}+{\frac {16}{15}}\,{\pi }^{4}-96\,\zeta  ( 3
 )  ) {n}^{3}
$$
$$
+ ( {\frac {15497}{9}}-240\,\ln  ( n
 ) -240\,\gamma+12\, ( \ln  ( n ) +\gamma
 ) ^{2}+1/6\, ( -916+144\,\ln  ( n ) +144\,\gamma
 ) {\pi }^{2}+8/5\,{\pi }^{4}-288\,\zeta  ( 3 ) 
 ) {n}^{2}
$$
$$
+ ( {\frac {11357}{9}}-310\,\ln  ( n ) 
-310\,\gamma+24\, ( \ln  ( n ) +\gamma ) ^{2}+1/6
\, ( -560+144\,\ln  ( n ) +144\,\gamma ) {\pi }^{
2}+{\frac {16}{15}}\,{\pi }^{4}-288\,\zeta  ( 3 )  ) n
$$
$$
-154\,\ln  ( n ) -154\,\gamma+12\, ( \ln  ( n
 ) +\gamma ) ^{2}+1/6\, ( -124+48\,\ln  ( n
 ) +48\,\gamma ) {\pi }^{2}+{\frac {4}{15}}\,{\pi }^{4}-96
\,\zeta  ( 3 ) \, + \, o(1) \quad .
$$
It follows that the limit of the scaled fourth moment (kurtosis) converges to
$$
{\frac{{\frac{2260}{9}}-28\,{\pi }^{2}+{\frac{4}{15}}\,{\pi }^{4}}{
 ( 7-2/3\,{\pi }^{2} ) ^{2}}}
\, = \,
4.1781156382698542\dots \quad .
$$

Results for higher moments, more precisely, up to the eighth moment, are also discovered and discussed by Shalosh B. Ekhad and Doron Zeilberger in [3].

Before this article, there are already human approaches to find the expectation and variance for the number of comparisons. Let $c_n = E[C_n]$. Since the pivot can be the $k$-th smallest element in the list $(k=1,2,\dots,n)$, we have the non-linear difference equation
$$
c_n = 
\frac{1}{n} \sum_{k=1}^{n} ((n-1)+ c_{k-1}+ c_{n-k} ) = (n-1) + \frac{1}{n} \sum_{k=1}^{n} (c_{k-1}+c_{n-k}) 
= (n-1) + \frac{2}{n} \sum_{k=1}^{n} \, c_{k-1},
$$
because the expected number of comparisons for the sublist before the pivot is $c_{k-1}$ and that for the sublist after the pivot is $c_{n-k}$. From this difference equation, complicated human-generated manipulatorics is needed to rigorously derive the closed form. For the variance, the calculation is much more complicated. For higher moments, we doubt that human approach is realistic. 

The experimental mathematics approach is more straightforward and more powerful. For the expectation, a list of data can be obtained through the difference equation and the initial condition. Then with an educated guess that $c_n$ is a polynomial of degree one in both $n$ and $H_n$, i.e.,
$$
c_n = a + bn + cH_n + dnH_n
$$
where $a,b,c,d$ are undetermined coefficients, we can solve for these coefficients by plugging sufficiently many $n$ and $c_n$ in this equation.

For higher moments, there is a similar difference equation for the probability generating function of $C_n$. With the probability generating function, a list of data of any fixed moment can be obtained. Then with another appropriate educated guess of the form of the higher moments, the explicit expression follows.  

In [3], it is already discussed that this experimental mathematics approach, which utilizes a difference equation to study the Quicksort algorithms, is actually rigorous by pointing out that this is a finite calculation and by referring to results in [16] and [17].

\section{Number of Swaps of 1-Pivot Quicksort}
The performance of Quicksort depends on the number of swaps and comparisons performed. In reality, a swap usually takes more computing resources than a comparison. The difficulty in studying the number of swaps is that the number of swaps depends on how we implement the Quicksort algorithm while the number of comparisons are the same despite the specific implementations. 

Since only the number of comparisons is considered in [3], the Quicksort model in [3] is that one picks the pivot randomly, compares each non-pivot element with the pivot and then places them in one of the two new lists $L_1$ and $L_2$ where the former contains all elements smaller than the pivot and the latter contains those greater than the pivot. Under this model there is no swap, but a lot of memory is needed. For convenience, let's call this model Variant Nulla. 

In this section, we consider the random variable, the number of swaps $X_n$, in different Quicksort variants. Some of them may not be efficient or widely used in industry; however, we treat them as an interesting problem and model in permutations and discrete mathematics. In the first subsection, we also demonstrate our experimental mathematics approaches step by step. 

\subsection{Variant I}
The first variant is that we always choose the first (or equivalently, the last) element in the list of length $n$ as the pivot, then we compare the other elements with the pivot. We compare the second element with the pivot first. If it is greater than the pivot, it stays where it is, otherwise we put it before the pivot and all the other elements including the pivot shift one position to the right. Though this is somewhat different than the ``traditional swap," we define this operation as a swap. Generally, every time we find an element smaller than the pivot, we put it on the pivot's current position and the indexes of the pivot and all elements on the right of pivot are incremented by one.

Hence, after $n-1$ comparisons and some number of swaps, the partition is achieved, i.e., all elements on the left of the pivot are smaller than the pivot and all elements on the right of the pivot are greater than the pivot. The difference between this variant and Variant Nulla is that this one does not need to create new lists so that it saves memory. 

Let $P_n(t)$ be the probability generating function for the number of swaps, i.e.,
$$
P_n(t) = \sum_{k=0}^{\infty}  P(X_n = k) \, t^k,
$$
where for only finitely many integers $k$, we have that $P(X_n = k)$ is nonzero.

We have the difference equation 
$$
P_n(t) = \frac{1}{n} \sum_{k=1}^n P_{k-1}(t) P_{n-k}(t) t^{k-1},
$$
with the initial condition $P_0(t)=P_1(t)=1$ because for any fixed $k \in \{1,2,\dots,n\}$, the probability that the pivot is the $k$-th smallest is $\frac{1}{n}$ and there are exactly $k-1$ swaps when the pivot is the $k$-th smallest.

The Maple procedure {\tt SwapPQs(n,t)} in the package {\tt Quicksort.txt} implements the recurrence of the probability generating function. 

Recall that the $r$-th moment is given in terms of the probability generating function
$$
E[X_n^r]= (t\frac{d}{dt})^r P_n(t) \, \vert_{t=1} .
$$
The moment about the mean
$$
m_r(n) := E[(X_n - c_n)^r] ,
$$
can be easily derived from the raw moments $\{ E[X_n^l] \,|\, 1 \leq l \leq r\}$, using the binomial theorem and
linearity of expectation. Another way to get the moment about the mean is by considering
$$
m_r(n)= (t\frac{d}{dt})^r (\frac{P_n(t)}{t^{c_n}}) \, \vert_{t=1} .
$$
Recall that 
$$
H_k(n) := \sum_{i=1}^n \frac{1}{i^k}.
$$
Our educated guess is that there exists a polynomial of $r+1$ variables $F_r(x_0, x_1, \dots, x_r)$ such that 
$$
m_r(n) = F_r(n, H_1(n), \dots, H_r(n)).
$$
With the Maple procedure {\tt QsMFn}, we can easily obtain the following theorems by just entering {\tt QsMFn(SwapPQs, t, n, Hn, r)} where $r$ represents the moment you are interested in. When $r=1$, it returns the explicit expression for its mean rather than the trivial ``first moment about the mean''.  

{\bf Theorem 3.1.1} The expectation of the number of swaps of Quicksort for a list of length $n$ under Variant I is
$$
E[X_n] = (n+1) H_n - 2n.
$$

{\bf Theorem 3.1.2} The variance of $X_n$ is 
$$
2n(n+2) - (n+1) H_1(n) - (n+1)^2 H_2(n).
$$

{\bf Theorem 3.1.3} The third moment about the mean of $X_n$ is 
$$
-\frac{9}{4} n (n+3)^2 + (4n+4) H_1(n) + 3(n+1)^2 H_2(n) + 2(n+1)^3 H_3(n).
$$

{\bf Theorem 3.1.4} The fourth moment about the mean of $X_n$ is 
$$
\frac{1}{18} n (335n^3 + 1568n^2 + 3067n + 2770) - 3(n+1)(4n^2+8n+9)H_1(n) +3(n+1)^2 H_1(n)^2
$$
$$
(-(12n^2+24n+19)(n+1)^2 + 6(n+1)^3 H_1(n) ) H_2(n) + 3(n+1)^4 H_2(n)^2
$$
$$
- 12(n+1)^3 H_3(n) - 6(n+1)^4 H_4(n).
$$

The explicit expressions for higher moments can be easily calculated automatically with the Maple procedure {\tt QsMFn} and the interested readers are encouraged to find those formulas on their own. 

\subsection{Variant II}
The second variant is similar to the first one. One tiny difference is that instead of choosing the first or last element as the pivot, the index of the pivot is chosen uniformly at random. For example, we choose the $i$-th element, which is the $k$-th smallest, as the pivot. Then we compare those non-pivot elements with the pivot. If $i\neq 1$, the first element will be compared with the pivot first. If it is smaller than the pivot, it stays there, otherwise it is moved to the end of the list and all other elements including the pivot shift one position to the left. After comparing all the left-side elements with the pivot, we look at those elements whose indexes are originally greater than $i$. If they are greater than the pivot, no swap occurs; otherwise move them to the pivot's current position and the indexes of the pivot and elements after the pivot but before the swapped element now are incremented by one. 

In this case, the recurrence of the probability generating function $P_n(t)$ is more complicated as a consequence of that the number of swaps given that $i$ and $k$ is known is still a random variable rather than a fixed number as the case in Variant I. 

Let $Q(n,k,i,t)$ be the probability generating function for such a random variable. In fact, given a random permutation in the permutation group $S_n$ and that the $i$-th element is $k$, the number of swaps equals to the number of elements which are before $k$ and greater than $k$ or after $k$ and smaller than $k$. Hence, if there are $j$ elements which are before $k$ and smaller than $k$, then there are $i-1-j$ elements which are before $k$ and greater than $k$ and there are $k-1-j$ elements which are after $k$ and smaller than $k$. So in this case the number of swaps is $i+k-2-2j$. 

Then we need to determine the range of $j$. Obviously it is at least 0. In total there are $k-1$ elements which are less than $k$, at most $n-i$ of them occurring after $k$, so $j \geq k-1-n+i$. As for the upper bound, since there are only $i-1$ elements before $k$, we have $j \leq i-1$. Evidently, $j \leq k-1$ as well. Therefore the range of $j$ is $[\,\max(k-1-n+i, 0),\, \min(i-1, k-1)\,]$.

As for the probability that there are exactly $j$ elements which are before $k$ and smaller than $k$, it equals to 
$$
\binom{i-1}{j} \prod_{s=0}^{j-1} \frac{k-1-s}{n-1-s} \prod_{s=0}^{i-j-2} \frac{n-k-s}{n-1-j-s}.
$$
Consequently, the probability generating function is
$$
Q(n,k,i,t) = \sum_{j=\max(k-1-n+i, 0)}^{\min(i-1, k-1)}  \binom{i-1}{j} \prod_{s=0}^{j-1} \frac{k-1-s}{n-1-s} \prod_{s=0}^{i-j-2} \frac{n-k-s}{n-1-j-s} t^{i+k-2-2j},
$$
which is implemented by the Maple procedure {\tt PerProb(n, k, i, t)}. For example, {\tt PerProb(9, 5, 5, t)} returns 
$$
\frac{1}{70} t^8 + \frac{8}{35} t^6 + \frac{18}{35} t^4 + \frac{8}{35} t^2 + \frac{1}{70}.
$$
We have the difference equation
$$
P_n(t) = \frac{1}{n^2} \sum_{k=1}^n  \sum_{i=1}^n P_{k-1}(t) P_{n-k}(t)  Q(n,k,i,t),
$$
with the initial condition $P_0(t)=P_1(t)=1$, which is implemented by the Maple procedure {\tt SwapPQ(n, t)}. The following theorems follow immediately.

{\bf Theorem 3.2.1} The expectation of the number of swaps of Quicksort for a list of length $n$ under Variant II is
$$
E[X_n] = (n+1) H_n - 2n.
$$

{\bf Theorem 3.2.2} The variance of $X_n$ is 
$$
\frac{1}{6} n(11n+17) - \frac{1}{3}(n+1) H_1(n) - (n+1)^2 H_2(n).
$$

{\bf Theorem 3.2.3} The third moment about the mean of $X_n$ is 
$$
-\frac{1}{6} n (14n^2+57n+73) + (2n+2) H_1(n) + (n+1)^2 H_2(n) + 2(n+1)^3 H_3(n).
$$

{\bf Theorem 3.2.4} The fourth moment about the mean of $X_n$ is 
$$
\frac{1}{90} n (1496n^3 + 5531n^2 + 8527n + 6922) - \frac{1}{15}(n+1)(55n^2+85n+173)H_1(n) +\frac{1}{3}(n+1)^2 H_1(n)^2
$$
$$
(-\frac{1}{3}(33n^2+51n+25)(n+1)^2 + 2(n+1)^3 H_1(n) ) H_2(n) + 3(n+1)^4 H_2(n)^2
$$
$$
-4(n+1)^3 H_3(n) - 6(n+1)^4 H_4(n).
$$

Higher moments can also be easily obtained by entering {\tt QsMFn(SwapPQ, t, n, Hn, r)} where $r$ represents the $r$-th moment you are interested in. 

Comparing with Variant I where the index of the pivot is fixed, we find that these two variants have the same expected number of swaps. However, the variance and actually all even moments of the second variant are smaller. Considering that the average performance is already $O(n \log n)$ which is not far from the best scenario, it is favorable that a Quicksort algorithm has smaller variance. In conclusion, for this model, a randomly-chosen-index pivot can improve the performance of the algorithm. 

\subsection{Variant III}
Next we'd like to study the most widely used in-place Quicksort. This algorithm is called Lomuto partition scheme, which is attributed to Nico Lomuto and popularized by Bentley in his book {\it Programming Pearls} and Cormen, {\it et al.} in their book {\it Introduction to Algorithms}. This scheme chooses a pivot that is typically the last element in the list. Two indexes, $i$, the insertion index, and $j$, the search index are maintained. The following is the pseudo code for this variant. 
{\obeylines
{\tt 
{\bf algorithm} quicksort(A, s, t) {\bf is}
   \quad {\bf if} s < t {\bf then}
     \qquad  p := partition(A, s, t)
     \qquad   quicksort(A, s, p - 1)
     \qquad quicksort(A, p + 1, t)
}
} 

{\obeylines
{\tt 
{\bf algorithm} partition(A, s, t) {\bf is}
  \quad   pivot := A[t]
  \quad  i := s
   \quad {\bf for} j := s {\bf to} t - 1 {\bf do}
    \qquad    {\bf if} A[j] < pivot {\bf then}
    \qquad \quad        swap A[i] with A[j]
        \qquad \quad    i := i + 1
  \quad  swap A[i] with A[t]
 \quad   {\bf return} i
}
}

From the algorithm we can see that when the pivot is the $k$-th smallest, there are $k-1$ elements smaller than the pivot. As a result, there are $k-1$ swaps in the {\tt if} statement of the algorithm {\tt partition}. Including the last swap outside the {\tt if} statement, there are $k$ total swaps. We have the difference equation for its probability generating function 
$$
P_n(t) = \frac{1}{n} \sum_{k=1}^n P_{k-1}(t) P_{n-k}(t) t^k
$$
with the initial condition $P_0(t)=P_1(t)=1$. 

{\bf Theorem 3.3.1} The expectation of the number of swaps of Quicksort for a list of length $n$ under Variant III
$$
E[X_n] = (n+1) H_n - \frac{4}{3} n -\frac{1}{3}.
$$

{\bf Theorem 3.3.2} The variance of the number of swaps of Quicksort for a list of length $n$ under Variant III
$$
var[X_n] = 2n^2 + \frac{187}{45} n +\frac{7}{45} - \frac{2}{3n} - (n^2+2n+1) H_2(n) - (n+1) H_1(n).
$$
Note that for this variant, and some other ones in the remainder of the paper, to find out its explicit expression of moments, we may need to modify our educated guess to a ``rational function'' of $n$ and $H_k(n)$ for some $k$ (see procedure {\tt QsMFnRat} and {\tt QsMFnRatG}). Moreover, sometimes when we solve the equations obtained by equalizing the guess with the empirical data, some initial terms should be disregarded since the increasing complexity of the definition of the Quicksort algorithms may lead to the ``singularity'' of the first several terms of moments. Usually, the higher the moment is, the more initial terms should be disregarded. 

\subsection{Variant IV}
In Variant III, every time when {\tt A[j] < pivot}, we swap {\tt A[i]} with {\tt A[j]}. However, it is a waste to swap them when $i=j$. If we modify the algorithm such that a swap is performed only when the indexes $i \neq j$, the expected cost will be reduced. Besides, if the pivot is actually the largest element, there is no need to {\tt swap A[i] with A[t]} in the partition algorithm. To popularize Maple in mathematical and scientific research, we attach Maple code for the partition part here, the {\tt ParIP} procedure in the package {\tt QuickSort.txt}, in which {\tt Swap(S, i, j)} is a subroutine to swap the $i$-th element with the $j$-th in a list $S$.
{\obeylines
{\tt 
ParIP:=proc(L) local pivot,i,j,n,S: 
n:=nops(L):
pivot:=L[n]:
S:=L:
i:=1:
for j from 1 to n-1 do
\quad  if S[j]<=pivot then
 \qquad   if i<>j then
  \quad \qquad    S:=Swap(S, i, j):
 \quad \qquad     i:=i+1:
  \qquad  else
  \quad \qquad     i:=i+1:
\qquad    fi:
\quad  fi:
od:
if i<>n then
\quad  return Swap(S, i, n), i:
else
\quad  return S, i:
fi:
end:
}
} 

{\bf Lemma 3.4.1} Let $Y_n(k)$ be the number of swaps needed in the first partition step in an in-place Quicksort without swapping the same index for a list $L$ of length $n$ when the pivot is the $k$-th smallest element, then
$$
Y_n(k) = 
\begin{cases} 
   |\, \{i \in [n]\,|\, L[i] \leq pivot \wedge \exists j<i, L[j] > pivot  \} \, |  & k<n \\
   0 & k=n
\end{cases}.
$$

\begin{proof}
When $k=n$, it is obvious that for each search index $j$, the condition {\tt S[j] <= pivot} is satisfied, hence the insertion index $i$ is always equal to $j$, which means there is no swap inside the loop. Since eventually $i=n$, there is also no need to swap the pivot with itself. So the number of swaps is $0$ in this case. 

When $k<n$, notice that the first time $i$ is smaller than $j$ is when we find the first element greater than the pivot. After that, $i$ will always be less than $j$, which implies that for each element smaller than the pivot and the pivot itself, one swap is performed. 
\end{proof}

The Maple procedure {\tt IPProb(n,k,t)} takes inputs $n,k$ as defined above and a symbol $t$, and outputs the probability generating function $Q(n,k,t)$ for the number of swaps in the first partition when the length of the list is $n$ and the last element, which is chosen as the pivot, is the $k$-th smallest.

When $k<n$, the probability that there are $s$ swaps is
$$
\binom{k-1}{k-s} \frac{(k-s)! (n-k) (n-k-2+s)! }{ (n-1)!} = \frac{n-k}{n-1} \frac{\binom{k-1}{k-s}}{\binom{n-2}{k-s}}.
$$
Therefore the probability generating function 
$$
Q(n,k,t) = \sum_{s=1}^k \frac{n-k}{n-1} \frac{\binom{k-1}{k-s}}{\binom{n-2}{k-s}} t^s.
$$
The difference equation for the probability generating function $P_n(t)$ of the total number of swaps follows immediately:
$$
P_n(t) = \frac{1}{n} \sum_{k=1}^n P_{k-1}(t) P_{n-k}(t) Q(n,k,t)
$$
with the initial condition $P_0(t)=P_1(t)=1$.

{\bf Theorem 3.4.2} The expectation of the number of swaps of Quicksort for a list of length $n$ under Variant IV
$$
E[X_n] = (n+2) H_n - \frac{5}{2} n -\frac{1}{2}.
$$

{\bf Theorem 3.4.3} The variance of the number of swaps of Quicksort for a list of length $n$ under Variant IV
$$
var[X_n] = 2n^{2}-{\frac {215}{12}}n+\frac{1}{12}+(11n+14)H_1(n) -(n^2-2n-2)H_2(n) - (2n+2) H_1(n)^2
$$

Comparing these results with {\bf Theorem 3.1.1} and {\bf Theorem 3.2.1}, it shows that Variant IV has better average performances, notwithstanding the ``broader'' definition of ``swap'' in the first two subsections. And of course, it is better than the in-place Quicksort which swaps the indexes even when they are the same. We fully believe that the additional cost to check whether the insertion and search indexes are the same is worthwhile.

\subsection{Variant V}
This variant might not be practical, but we find that it is interesting as a combinatorial model. As is well-known, if a middle-most element is chosen as a pivot, the Quicksort algorithm will have better performance than average in this case. Hence if additional information is available so that the probability distribution of chosen pivots is no longer a uniform distribution but something Gaussian-like, it is to our advantage. 

Assume that the list is a permutation of $[n]$ and we are trying to sort it, pretending that we do not know the sorted list must be $[1,2, \dots, n]$. Now the rule is that we choose the first and last number in the list, look at the numbers and choose the one which is closer to the middle. If the two numbers are equally close to the middle, then choose one at random.

Without loss of generality, we can always assume that the last element in the list is chosen as the pivot; otherwise we just need to run the algorithm in the last subsection ``in reverse'', putting both the insertion and search indexes on the last element and letting larger elements be swapped to the left side of the pivot, etc. So the only difference with Variant IV is the probability distribution of $k$, which is no longer $\frac{1}{n}$ for each $k \in [n].$

Considering symmetry, $Pr^{(n)}(pivot = k) = Pr^{(n)}(pivot = n+1-k)$, so we only need to consider $1 \leq k \leq (n+1)/2$. When $n$ is even, let $n=2m$. Then $Pr^{(n)}(pivot = k) = \frac{4k-3}{(2m-1)2m}$ . When $n$ is odd, let $n = 2m-1$, then $Pr^{(n)}(pivot = k) = \frac{4k-3}{(2m-1)(2m-2)}$ when $k<m$ and $Pr^{(n)}(pivot = m) = \frac{2}{2m-1}$.

With this minor modification, the difference equation for $P_n(t)$ follows.
$$
P_n(t) =  \sum_{k=1}^n P_{k-1}(t) P_{n-k}(t) Q(n,k,t) Pr^{(n)}(pivot = k)
$$
with the initial condition $P_0(t)=P_1(t)=1$.

Though an explicit expression seems difficult in this case, we can still analyze the performance of the algorithm by evaluating its expected number of swaps. By exploiting the Maple procedure {\tt MomFn(f,t,m,N)}, which inputs a function name $f$, a symbol $t$, the order of the moment $m$ and the upper bound of the length of the list $N$ and outputs a list of $m$-th moments for the Quicksort described by the probability generating function $f$ of lists of length $1,2, \dots, N$, we find that Variant V has better average performance than Variant IV when $n$ is large enough. For example, {\tt MomFn(PQIP, t, 1, 20)} returns
$$
[0,\frac{1}{2},\frac{7}{6},2,{\frac {179}{60}},{\frac {41}{10}},{\frac {747}{140}},{
\frac {187}{28}},{\frac {20459}{2520}},{\frac {1013}{105}},{\frac {
312083}{27720}},{\frac {25631}{1980}},{\frac {353201}{24024}},{\frac {
1488737}{90090}},
$$
$$
{\frac {6634189}{360360}},{\frac {814939}{40040}},{
\frac {273855917}{12252240}},{\frac {4983019}{204204}},{\frac {
97930039}{3695120}},{\frac {20210819}{705432}}],
$$
and {\tt MomFn(PQIPk, t, 1, 20)} returns 
$$
[0,\frac{1}{2},\frac{4}{3},{\frac {20}{9}},{\frac {155}{48}},{\frac {1957}{450}},{
\frac {2341}{420}},{\frac {4055}{588}},{\frac {55829}{6720}},{\frac {
794}{81}},{\frac {630547}{55440}},{\frac {170095}{13068}},{\frac {
12735487}{864864}},
$$
$$
{\frac {3864281}{234234}},{\frac {2521865}{137592}}
,{\frac {36424327}{1801800}},{\frac {4343228489}{196035840}},{\frac {
107768347}{4463316}},{\frac {15673532207}{598609440}},{\frac {
1136599735}{40209624}}].
$$
We notice that for $n \geq 14$, Variant V consistently has better average performance. From this result we can conclude that it is worth choosing a pivot from two candidates since the gains of efficiency are far more than its cost. 

Moreover, we can obtain a recurrence for the expected number $x_n$ of the random variable $X_n$. The {\tt Findrec(f,n,N,MaxC)} procedure in the Maple package {\tt Findrec.txt} inputs a list of data $L$, two symbols $n$ and $N$, where $n$ is the discrete variable, and $N$ is the shift operator, and $MaxC$ which is the maximum degree plus order. {\tt Findrec(MomFn(PQIPk, t, 1, 80), n, N, 11)} returns

\begin{figure}[h!]
  \includegraphics[width=\textwidth]{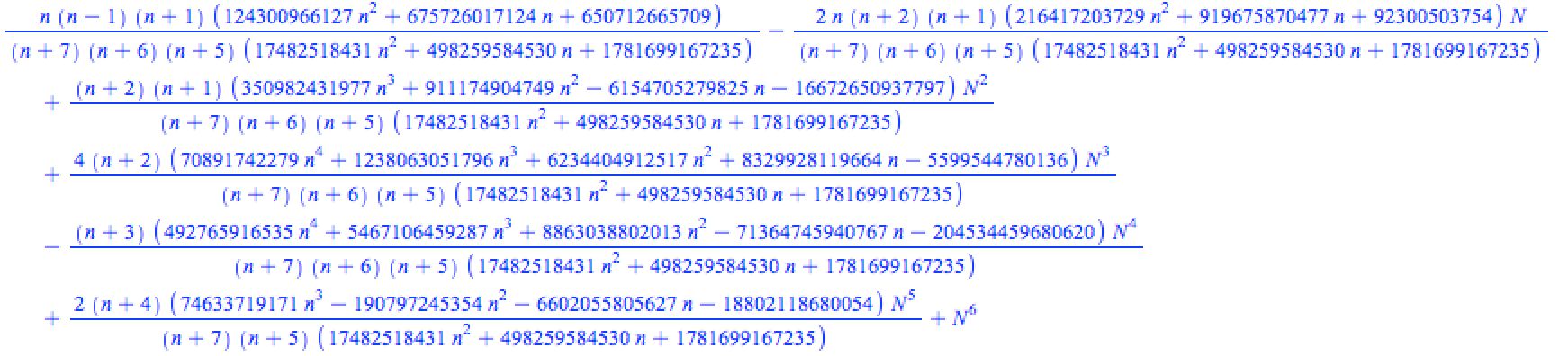}
  \caption{The Recurrence Relation for the Expected Number of Swaps}
\end{figure}

As previously mentioned, $N$ is the shift operator. Since this formula is too long, to see its detailed interpretation, please look at {\bf Theorem 4.2.1} as reference. 

We can also look at more elements to choose the middle-most one as the pivot. In case that we do not want to store so much information, 
some other variants involving a ``look twice'' idea could be that if the first selected element is within some satisfactory interval, e.g., $[\frac{n}{4}, \frac{3n}{4}]$ for a permutation of $n$, then it is chosen
as the pivot. Otherwise  we choose a second element as the pivot without storing information about the first one. It is also likely to improve the algorithm with ``multiple looks'' to choose the pivot and the requirement to choose the current element as the pivot without continuing to look at the next one could vary and ideally the requirement should be more relaxed as the number of ``looks'' increases. We also refer the readers to [13] where P. Kirschenhofer, H. Prodinger and C. Martinez chose the median of a random sample of three elements as the pivot and obtained explicit expressions for this variant's performance with methods of hypergeometric differential equations. In general, it is ideal to have a probability distribution of pivots where an element closer to the median is more likely to be chosen.

\section{Explorations for Multi-Pivot Quicksort}
With the same general idea as the 1-pivot Quicksort, it is natural to think about ``what if we choose more pivots." In $k$-pivot Quicksort, $k$ indexes are chosen and the correspondent elements become pivots. The $k$ pivots need to be sorted and then the other elements will be partitioned into one of the $k+1$ sublists. Compared to 1-pivot Quicksort, multi-pivot Quicksort is much more complicated because we need to determine how to sort the pivots, how to allocate each non-pivot element to the sublist they belong to and how to sort a sublist containing less than $k$ elements. Therefore, there are a few multi-pivot Quicksort variants. We refer the reader to [9] for other versions of multi-pivot. When $k$ is large, it seems difficult to have an in-place version, so we mainly consider the random variable, the number of comparisons $C_n$, in this section since a swap might be difficult to define in this case. 

\subsection{Dual-Pivot Quicksort}
Let's start from the simplest multi-pivot Quicksort: dual-pivot. The model for dual-pivot Quicksort is that two pivots $p_1$ and $p_2$ are randomly chosen. After one comparison, they are sorted, say $p_1<p_2$. Non-pivot elements should be partitioned into one of the three sublists $L_1, L_2$ and $L_3$. $L_1$ contains elements smaller than $p_1$, $L_2$ contains elements between $p_1$ and $p_2$ while $L_3$ contains elements greater than $p_2$. Each non-pivot element is compared with $p_1$ first. If it is smaller than $p_1$, we are done. Otherwise it is compared with $p_2$ to determine whether it should go to $L_2$ or $L_3$.

Given that the list contains $n$ distinct elements and the two pivots are the $i$-th and $j$-th smallest element $(i<j)$, we need one comparison to sort the pivot and $i-1+2(n-i-1) = 2n-i-3$ comparisons to distribute non-pivot elements to the sublists. Hence the total number of comparison is $2n-i-2$ and the difference equation for the probability generating function $P_n(t)$ of the total number of comparisons $C_n$ of dual-pivot Quicksort is
$$
P_n(t) = \frac{1}{\binom{n}{2}} \sum_{j=2}^n \sum_{i=1}^{j-1}  P_{i-1}(t) P_{j-i-1}(t) P_{n-j}(t) t^{2n-i-2}
$$
with the initial condition $P_0(t)=P_1(t)=1$ and $P_2(t) = t$.

The above recurrence is implemented by the procedure {\tt PQc2}. With the aforementioned Maple procedure {\tt QsMFn} it is easy to get the following theorems.

{\bf Theorem 4.1.1} The expectation of the number of comparisons in dual-pivot Quicksort algorithms is
$$
E[C_n]=2(n+1) H_n - 4n .
$$

{\bf Theorem 4.1.2} The variance of the number of comparisons in dual-pivot Quicksort algorithms is
$$
var[C_n] = n ( 7\,n+13 )  \, - \, 2\,(n+1)\, H_{{1}} ( n )  -4\, ( n+1 ) ^{2}H_{{2}} ( n )  .
$$

{\bf Theorem 4.1.3}  The third moment about the mean of $C_n$ is 
$$
-n ( 19\,{n}^{2}+81\,n+104 ) +H_{{1}} ( n ) 
 ( 14\,n+14 ) +12\, ( n+1 ) ^{2}H_{{2}} ( n
 ) +16\, ( n+1 ) ^{3}H_{{3}} ( n )  .
$$

{\bf Theorem 4.1.4} The fourth moment about the mean of $C_n$ is 
$$
\frac{1}{9}\,n ( 2260\,{n}^{3}+9658\,{n}^{2}+15497\,n+11357 ) -2\,
 ( n+1 )  ( 42\,{n}^{2}+78\,n+77 ) H_{{1}}
 ( n ) 
$$
$$
+12\, ( n+1 ) ^{2} ( H_{{1}} ( n
 )  ) ^{2}+ ( -4\, ( 42\,{n}^{2}+78\,n+31
 )  ( n+1 ) ^{2}+48\, ( n+1 ) ^{3}H_{{1}}
 ( n )  ) H_{{2}} ( n ) 
$$
$$
+48\, ( n+1
 ) ^{4} ( H_{{2}} ( n )  ) ^{2}-96\,
 ( n+1 ) ^{3}H_{{3}} ( n ) -96\, ( n+1
 ) ^{4}H_{{4}} ( n ) .
$$

As usual, any higher moment can be easily obtained with a powerful silicon servant. Careful readers may notice that the above four theorems are exactly the same as the ones in Section 2. It is natural to ask whether they indeed have the same probability distribution. The answer is yes. In Section 4.1.2 of [8] the author gives a sketch of proof showing that 1-pivot and dual-pivot Quicksorts's numbers of comparisons satisfy the same recurrence relation and initial condition. From an experimental mathematical point of view, a semi-rigorous proof obtained by checking sufficiently many special cases is good enough for us. For example, the first 10 probability generating function, $P_n(t)$ for $1 \leq n \leq 10$ can be calculated in a nanosecond by entering {\tt [seq(PQc2(i, t), i = 1..10)]} and we have
$$
P_1(t) = 1,
$$
$$
P_2(t) = t,
$$
$$
P_3(t) = \frac{2}{3} t^3 + \frac{1}{3} t^2,
$$
$$
P_4(t) = \frac{1}{3}{t}^{6}+\frac{1}{6}{t}^{5}+\frac{1}{2}{t}^{4},
$$
$$
P_5(t) = \frac{2}{15}{t}^{10}+\frac{1}{15}{t}^{9}+\frac{1}{5}{t}^{8}+\frac{4}{15} t^7 +\frac{1}{3} t^6,
$$
$$
\vdots
$$
which are exactly the same as the probability generating function for the number of comparisons of 1-pivot Quicksort. 

In conclusion, in terms of probability distribution of the number of comparisons, the dual-pivot Quicksort does not appear to be better than the ordinary 1-pivot Quicksort. 

As for the random variable $X_n$, the number of swaps, it depends on the specific implementation of the algorithm and the definition of a ``swap." As a toy model, we do an analogue of Section 3.1. Assume the list is a permutation of $[n]$. The first and last elements are chosen as the pivot. Let's say they are $i$ and $j$. If $i>j$ then we swap them and still call the smaller pivot $i$. For each element less than $i$, we move it to the left of $i$, and for each element greater than $j$, we move it to the right of $j$ and call this kind of operations a swap. 

The difference equation for the probability generating function of $X_n$ is
$$
P_n(t) = \frac{1}{\binom{n}{2}} (\frac{1}{2} + \frac{1}{2} t) \sum_{j=2}^n \sum_{i=1}^{j-1} P_{i-1}(t) P_{j-i-1}(t) P_{n-j}(t) t^{n-1+i-j}
$$
with the initial conditions $P_0(t) = P_1(t) =1$ and $P_2(t) =\frac{1}{2} + \frac{1}{2} t. $

{\bf Theorem 4.1.5} The expectation of the number of swaps in the above dual-pivot Quicksort variant is
$$
E[X_n]=\frac{4}{5} (n+1) H_n - \frac{39}{25} n -\frac{1}{100}.
$$
Note that this result is better than those in Sections 3.1 and 3.2.

\subsection{Three-Pivot Quicksort}
As mentioned at the beginning of this section, to define a 3-pivot Quicksort, we need to define 1) how to sort the pivots, 2) how to partition the list and 3) how to sort a list or sublist containing less than 3 pivots. Since this paper is to study Quicksort, we choose 1-pivot Quicksort for 1) and 3). For 2), it seems that the binary search is a good option since for each non-pivot element exactly 2 comparisons with the pivots are needed. 

The Maple procedure {\tt PQs(n,t)} outputs the probability generating function for 1-pivot Quicksort of a list of length $n$. Hence during the process of sorting the pivots and partitioning the list, the probability generating function of the number of comparisons is $PQs(3,t) t^{2n-6}$, which equals to
$$
(\frac{2}{3} t^3 + \frac{1}{3} t^2) t^{2n-6} = \frac{2}{3} t^{2n-3} + \frac{1}{3} t^{2n-4}.
$$
So the difference equation for the probability generating function $P_n(t)$ of the total number of comparisons for 3-pivot Quicksort of a list of length $n$ is
$$
P_n(t) = \frac{1}{\binom{n}{3}} \sum_{k=3}^n \sum_{j=2}^{k-1} \sum_{i=1}^{j-1} P_{i-1}(t) P_{j-i-1}(t) P_{k-j-1}(t) P_{n-k}(t) (\frac{2}{3} t^{2n-3} + \frac{1}{3} t^{2n-4} )
$$
with initial conditions $P_0(t) = P_1(t) =1, P_2(t) = t$ and $P_3(t) =\frac{2}{3} t^3 + \frac{1}{3} t^2 $. This recurrence is implemented by the procedure {\tt PQd3}.

The explicit expression seems to be difficult to obtain in this case, but numerical tests imply that 3-pivot Quicksort has better performances than dual-pivot, and of course 1-pivot since it is indeed equivalent to dual-pivot.  

By exploiting the Maple procedure {\tt MomFn(f,t,m,N)} again, we can compare the expectations of different Quicksort variants. 

For instance, {\tt MomFn(PQc2, t, 1, 20)} returns 
$$
[0,1,\frac{8}{3},{\frac {29}{6}},{\frac {37}{5}},{\frac {103}{10}},{\frac {472}{35}},{\frac {2369}{140}},{\frac {2593}{126}},{\frac {30791}{1260}},{
\frac {32891}{1155}},{\frac {452993}{13860}},{\frac {476753}{12870}},{
\frac {499061}{12012}},
$$
$$
{\frac {2080328}{45045}},{\frac {18358463}{
360360}},{\frac {18999103}{340340}},{\frac {124184839}{2042040}},{
\frac {127860511}{1939938}},{\frac {26274175}{369512}}],
$$
and {\tt MomFn(PQd3, t, 1, 20)} returns
$$
[0,1,\frac{8}{3},\frac{14}{3},{\frac {106}{15}},{\frac {49}{5}},{\frac {64}{5}},{\frac {561}{35}},{\frac {1226}{63}},{\frac {5192}{225}},{\frac {465316}{17325}},{\frac {533509}{17325}},{\frac {714008}{20475}},{\frac {
61615768}{1576575}},
$$
$$
{\frac {342234824}{7882875}},{\frac {754600981}{
15765750}},{\frac {1404956027}{26801775}},{\frac {15298397599}{
268017750}},{\frac {31489234438}{509233725}},{\frac {1697926310039}{
25461686250}}].
$$
We notice that for each fixed $n>3$, 3-pivot Quicksort's average performance is better than 2-pivot and 1-pivot. This numerical test is also possible for all previous Quicksort variants but seems unnecessary when the explicit expressions are easily accessible. 

As in Section 3.5, a difference equation of the expected number of comparisons can be obtained. {\tt Findrec(MomFn(PQd3, t, 1, 40),n,N,8)} returns 
$$
{\frac { \left( 3\,n+4 \right)  \left( {n}^{2}-5\,n+12 \right) }{
 \left( n+4 \right)  \left( n+3 \right)  \left( 3\,n+1 \right) }}-{
\frac { \left( 12\,{n}^{4}+13\,{n}^{3}-12\,{n}^{2}+59\,n+24 \right) N
}{ \left( 3\,n+1 \right)  \left( n+4 \right)  \left( n+3 \right) 
 \left( n+2 \right) }}
$$
$$
 +3\,{\frac { \left( n+1 \right)  \left( 6\,n+5
 \right) n{N}^{2}}{ \left( n+4 \right)  \left( n+3 \right)  \left( 3\,
n+1 \right) }}-{\frac { \left( n+1 \right)  \left( 12\,n+7 \right) {N}
^{3}}{ \left( n+4 \right)  \left( 3\,n+1 \right) }}+{N}^{4},
$$
which leads to the following theorem.

{\bf Theorem 4.2.1} The expected number of comparisons $C_n$ of 3-pivot Quicksort for a list of length $n$ satisfies the following difference equation:
$$
C_{n+4} = {\frac { \left( n+1 \right)  \left( 12\,n+7 \right) }{ \left( n+4 \right)  \left( 3\,n+1 \right) }} C_{n+3} - 3\,{\frac { \left( n+1 \right)  \left( 6\,n+5 \right) n}{ \left( n+4 \right)  \left( n+3 \right)  \left( 3\,n+1 \right) }} C_{n+2}
$$
$$
+{\frac { \left( 12\,{n}^{4}+13\,{n}^{3}-12\,{n}^{2}+59\,n+24 \right) 
}{ \left( 3\,n+1 \right)  \left( n+4 \right)  \left( n+3 \right) 
 \left( n+2 \right) }} C_{n+1} - {\frac { \left( 3\,n+4 \right)  \left( {n}^{2}-5\,n+12 \right) }{
 \left( n+4 \right)  \left( n+3 \right)  \left( 3\,n+1 \right) }} C_{n}.
$$
The difference equations for higher moments are also obtainable, but a long enough list of data is needed. 

\subsection{$k$-Pivot Quicksort}
More generally, $k$-pivot Quicksort can be considered with the convention that 1) the $k$ pivots are sorted with 1-pivot Quicksort, 2) binary search is used to partition the list into $k+1$ sublists, 3) we use 1-pivot Quicksort to sort lists with length less than $k$.

In the package {\tt QuickSort.txt} the procedure {\tt PQck(n, t, k)} outputs the probability generating function for the number of comparisons of $k$-pivot Quicksort where each element is compared with pivots in a linearly increasing order. Obviously this is not efficient when $k$ is large. However, the problem for binary search is that when $k \neq 2^i-1$ for some $i$, it is hard to get an expression for the number of comparisons in the binary search step since the number highly depends on the specific implementation where some boundary cases may vary and the floor and ceiling functions will be involved, which leads to an increasing difficulty to find the explicit expressions for moments. 

There is a procedure {\tt QsBC(L, k)} which inputs a list of distinct numbers $L$ and an integer $k$ representing the number of pivots and outputs the sorted list and the total number of comparisons. For convenience of Monte Carlo experiments, we use {\tt MCQsBC(n, k, T)} where $n$ is the length of the list, $k$ is the number of pivots and $T$ is the number of times we repeat the experiments. Because of the limit of computing resources, we only test for $k=3,4,5,6$ and $n=10, 20 ,30 ,40 ,50.$
$$
{\tt [seq(MCQsBC(10*i, 3, 100), i = 1 .. 5)]} = [22.95, 65.75, 118.71, 178.28, 239.45],
$$

$$
{\tt [seq(MCQsBC(10*i, 4, 100), i = 1 .. 5)]} = [23.78, 67.77, 120.91, 180.35, 251.19],
$$

$$
{\tt [seq(MCQsBC(10*i, 5, 100), i = 1 .. 5)]} = [23.54, 65.74, 119.59, 178.36, 241.03],
$$

$$
{\tt [seq(MCQsBC(10*i, 6, 100), i = 1 .. 5)]} = [23.14, 66.22, 120.07, 176.43, 236.46],
$$

Our observation is that for large enough $n$, the more pivots we use, the less comparisons are needed. However, when $k$ is too close to $n$, the increase of pivots may lead to inefficiency. 

\section{Limiting Distribution}
The main purpose of this paper is to find explicit expressions for the moments of the number of swaps or comparisons of some variants of Quicksort, to compare their performances and to explore more efficient Quicksort algorithms. However, it is also of interest to find more moments for large $n$ and calculate their floating number approximation of the scaled limiting distribution. 

As mentioned in [3], if we are only interested in the first few moments, then it is wasteful to compute the full probability generating function $P_n(t).$
Let $t=1+w$ and use the fact that
$$
P_n(1+w)  = \sum_{r=0}^{\infty} \frac{f_r(n)}{r!} w^r
$$
where $f_r(n)$ are the factorial moments. The straight moments $E[X_n^r]$, and the moments-about-the-mean,
$m_r(n)$ follow immediately. 

As a case study, let's use Variant IV from Section 3.4 as an example. The difference equation is 
$$
P_n(1+w)  = \frac{1}{n} \sum_{k=1}^{n} P_{k-1}(1+w) P_{n-k}(1+w) Q(n,k,1+w), 
$$
where $Q$ is as defined in Section 3.4.

Since only the first several factorial moments are considered, in each step truncation is performed and only the first several coefficients in $w$ is kept. With this method we can get more moments in a fixed time. The procedure {\tt TrunIP} implements the truncated factorial generating function. 

With the closed-form expressions for both the expectation, $c_n$, and the variance $m_2(n):=var(X_n)$, the scaled random variable $Z_n$ is defined as follows.
$$
Z_n := \frac{X_n -c_n}{\sqrt{m_2(n)}} .
$$

We are interested in the floating point approximations of the limiting distribution $\lim_{n \rightarrow \infty} Z_n$. Of course its expectation is $0$ and its variance is $1$.

For instance, if we'd like to know the moments up to order 10, {\tt TrunIP(100, z, 10)} returns
$$
1+{\frac {7617634712836831344646726224164628686543}{
27341323619495089084130905464828354336}} z
$$
$$
+{\frac {
1169146867836246319480317311960440606057785761234433183813484643}{29517287662514914280390084303910684938635848245569645536000}} z^2
$$
$$
+{
\frac {
58024172013839694810625346567417182291098218339356411215067112605982034521
}{
15125688216961909953814450921738787993181911018772132633289881600000}} z^3
$$
$$
+ \dots \quad.
$$
For $1 \leq r \leq 10$, the coefficient of $z^r$ times $r!$ is the $r$-th factorial moment. By 
$$
E[X^r] = \sum_{j=0}^r {r\brace j} E[(X)_j]
$$
where the curly braces denote Stirling numbers of the second kind, we can get the raw moments. And with the procedure {\tt MtoA} a sequence of raw moments are transformed to moments about the mean. Divided by $m_2(n)^{\frac{r}{2}}$, the 3rd through 10th moments
in floating point approximations are
$$
[0.7810052982, 3.942047050, 9.146681877, 37.12169647, 137.7143092,
$$
$$
613.5286860, 2872.409923, 14709.75560].
$$
The same technique can be applied to other variants of Quicksort in this paper and we leave this to interested readers.

\section{Future Work and Final Remarks}
In such a rich and active research area as Quicksort, there are still several things we could think about to improve the algorithms' performances. Just to name a few, in 2-pivot Quicksort when we compare non-pivot elements with the pivots to determine which sublist they belong to, if the history is tracked, we might be able to use the history to determine which pivot to compare with first for the next element. The optimal strategy would vary with the additional information about the range of the numbers or the relative ranking of the two pivots among all the elements. 

As for $k$-pivot Quicksort, our naive approach only distinguishes two cases: whether the currently to-be-sorted list has length less than $k$ or not. If the length is less than $k$, we use 1-pivot Quicksort; otherwise we still choose $k$ pivots. However, we might be able to improve the performance if the number of pivots varies according to the length of the to-be-sorted list or sublist. Let's say there is a function $g(n)$, where $n$ is the length of the list. So we pick $g(n)$ pivots at the beginning. After we obtain the $g(n)+1$ sublists with length $n_i, 1 \leq i \leq g(n)+1$, for each one of them, we choose $g(n_i)$ pivots. It would be interesting whether we can find an optimal $g$ in terms of its average performance. Additionally, when $k$ is large, it might make sense to use multi-pivot Quicksort to sort the $k$ pivots as well. 

Of course, it is also interesting to study the explicit expressions of the numbers of swaps in multi-pivot Quicksort. But it appears to be dependent on the specific implementation of the algorithm so it is of significance to look for variants which save time and space complexity. 

The main results of this paper are those explicit expressions of moments and difference equations for either the number of comparisons or the number of swaps of various Quicksort variants. Though all of their asymptotics are $O(n \log n)$, the constant before this term varies a lot and some comparisons of these variants are also discussed. When there is difficulty getting the explicit expressions, numerical tests and Monte Carlo experiments are performed. We also have a demonstration on how to get more moments and find the numeric approximation of the scaled limiting distribution. 

Nevertheless, more important than those results is the illustration of a methodology of experimental mathematics. From ansatzes and sufficient data we have an alternative way to obtain results that otherwise might be extremely difficult or even impossible to get via traditional human approaches to algorithm analysis.

\hfill\break

{\bf Acknowledgment}
I would like to express my special thanks of gratitude to my advisor Doron Zeilberger for his mentoring and guidance. I am also grateful to Lun Zhang for interesting conversations and useful remarks. Many thanks are due to Vasileios Iliopoulos for his helpful clarifications and suggestions. 

\hfill\break

{\bf References}

[1] Michael Cramer, {\it A note concerning the limit distribution of the quicksort algorithm},
Informatique The\'eriques et Applications, {\bf 30}, 195-207, 1996.

[2] Marianne Durand, {\it Asymptotic analysis of an optimized Quicksort algorithm}, Inform. Proc. Lett., {\bf 85} (2): 73-77, 2003

[3] Shalosh B. Ekhad and Doron Zeilberger,
{\it A Detailed Analysis of Quicksort Running Time},
The Personal Journal of Shalosh B. Ekhad and Doron Zeilberger, \hfill\break
{\tt http://www.math.rutgers.edu/~zeilberg/pj.html}  . \hfill\break
Direct url: \\{\tt http://sites.math.rutgers.edu/~zeilberg/mamarim/mamarimhtml/qsort.html}. 
Also in: {\tt https://arxiv.org/abs/1903.03708}  .

[4] Shalosh B. Ekhad and Doron Zeilberger,
{\it Explicit Expressions for the Variance and Higher Moments of the Size of a Simultaneous Core Partition and its Limiting Distribution},
The Personal Journal of Shalosh B. Ekhad and Doron Zeilberger, \hfill\break
{\tt http://www.math.rutgers.edu/~zeilberg/pj.html}  . \hfill\break
Direct url: \\{\tt http://www.math.rutgers.edu/~zeilberg/mamarim/mamarimhtml/stcore.html}. 
Also in: {\tt https://arxiv.org/abs/1508.07637}  .

[5] Ronald L. Graham, Donald E. Knuth, and Oren Patashnik, {\it ``Concrete Mathematics''},  Addison-Wesley, 1989.

[6] P. Hennequin, {\it Combinatorial analysis of quicksort algorithm}, RAIRO Theor. Inform. Appl., {\bf 23} (3): 317-333, 1989

[7] C.A.R. Hoare, {\it Quicksort}, The Computer Journal, {\bf 5}:1, 10-15, 1962.

[8] Vasileios Iliopoulos, {\it The Quicksort algorithm and related topics}, \hfill\break
{\tt https://arxiv.org/abs/1503.02504}.

[9] Vasileios Iliopoulos, {\it A note on multipivot Quicksort}, Journal of Information and Optimization Sciences, {\bf 39}:5, 1139-1147, 2018.

[10] Vasileios Iliopoulos and David B. Penman, {\it Dual pivot Quicksort}, Discrete Mathematics, Algorithms and Applications, Vol. 04, No. 03, 1250041, 2012.

[11] Donald E. Knuth, {\it The Art of Computer Programming, Volume 3: Sorting and Searching}, Addison-Wesley, 1973.

[12] Manuel Kauers and Peter Paule, {\it The Concrete Tetrahedron}, Springer, 2011.

[13] P. Kirschenhofer,  H. Prodinger,  C. Martinez, {\it Analysis of Hoare's FIND algorithm with Median-of-three partition}, Random Structures and Algorithms, {\bf 10}:1-2, 143-156, 1997.

[14] Charles Knessl and Wojciech Szpankowski, {\it Quicksort algorithm again revisited},
Discrete Mathematics and Theoretical Computer Science, {\bf 3}, 43-64, 1999.

[15] Alois Panholzer, {\it Analysis of multiple quickselect variants}, Theor. Comput. Sci., {\bf 302} (1-3): 45-91, 2003

[16] Carsten Schneider, {\it The Summation package Sigma}, A Mathematica package available from \hfill\break
{\tt https://www3.risc.jku.at/research/combinat/software/Sigma/index.php} 

[17] Carsten Schneider, {\it  Symbolic Summation Assists Combinatorics}, Sem.Lothar.Combin. {\bf 56}(2007),Article B56b (36 pages). \hfill\break
{\tt https://www3.risc.jku.at/research/combinat/software/Sigma/pub/SLC06.pdf}  

[18] Wikipedia, {\it , Quicksort}, \hfill\break
{\tt https://en.wikipedia.org/wiki/Quicksort} 

[19] Doron Zeilberger, {\it A Holonomic Systems Approach To Special Functions},
J. Computational and Applied Math {\bf 32} (1990), 321-368, \hfill\break
{\tt http://www.math.rutgers.edu/\~{}zeilberg/mamarim/mamarimPDF/holonomic.pdf}.

[20] Doron Zeilberger, {\it The Automatic Central Limit Theorems Generator (and Much More!)},
in: ``Advances in Combinatorial Mathematics: Proceedings of the Waterloo Workshop in 
Computer Algebra 2008 in honor of Georgy P. Egorychev'', chapter 8, pp. 165-174, (I.Kotsireas, E.Zima, eds. Springer Verlag, 2009),
\hfill\break
{\tt http://www.math.rutgers.edu/\~{}zeilberg/mamarim/mamarimhtml/georgy.html}.

[21]  Doron Zeilberger, {\it HISTABRUT: A Maple Package for Symbol-Crunching in Probability theory},
the Personal Journal of Shalosh B. Ekhad and Doron Zeilberger, posted Aug. 25, 2010,
\hfill\break
{\tt http://www.math.rutgers.edu/\~{}zeilberg/mamarim/mamarimhtml/histabrut.html}.

[22]  Doron Zeilberger, {\it Symbolic Moment Calculus I.: Foundations and Permutation Pattern Statistics},
Annals of Combinatorics {\bf 8}, 369-378, 2004 Available from:
\hfill\break
{\tt http://www.math.rutgers.edu/\~{}zeilberg/mamarim/mamarimhtml/smcI.html}.

\bigskip
\hrule
\bigskip
Yukun Yao, Department of Mathematics, Rutgers University (New Brunswick), Hill Center-Busch Campus, 110 Frelinghuysen
Rd., Piscataway, NJ 08854-8019, USA. \hfill\break
Email: {\tt yao at math dot rutgers dot edu}    .

\end{document}